\documentstyle[epsfig,twoside]{article}

\catcode`\@=11
\long\def\@makefntext#1{
\protect\noindent \hbox to 3.2pt {\hskip-.9pt  
$^{{\eightrm\@thefnmark}}$\hfil}#1\hfill}		%CAN BE USED 

\def\@makefnmark{\hbox to 0pt{$^{\@thefnmark}$\hss}}	%ORIGINAL 
	
\def\ps@myheadings{\let\@mkboth\@gobbletwo
\def\@oddhead{\hbox{}
\rightmark\hfil\eightrm\thepage}   
\def\@oddfoot{}\def\@evenhead{\eightrm\thepage\hfil
\leftmark\hbox{}}\def\@evenfoot{}
\def\sectionmark##1{}\def\subsectionmark##1{}}

\oddsidemargin=\evensidemargin
\newcounter{sectionc}\newcounter{subsectionc}\newcounter{subsubsectionc}
\renewcommand{\section}[1] {\vspace{12pt}\addtocounter{sectionc}{1} 
\setcounter{subsectionc}{0}\setcounter{subsubsectionc}{0}\noindent 
	{\tenbf\thesectionc. #1}\par\vspace{5pt}}
\renewcommand{\subsection}[1] {\vspace{12pt}\addtocounter{subsectionc}{1} 
	\setcounter{subsubsectionc}{0}\noindent 
	{\bf\thesectionc.\thesubsectionc. {\kern1pt \bfit #1}}\par\vspace{5pt}}
\renewcommand{\subsubsection}[1] {\vspace{12pt}\addtocounter{subsubsectionc}{1}
	\noindent{\tenrm\thesectionc.\thesubsectionc.\thesubsubsectionc.
	{\kern1pt \tenit #1}}\par\vspace{5pt}}
\newcommand{\nonumsection}[1] {\vspace{12pt}\noindent{\tenbf #1}
	\par\vspace{5pt}}

\newcounter{appendixc}
\newcounter{subappendixc}[appendixc]
\newcounter{subsubappendixc}[subappendixc]
\renewcommand{\thesubappendixc}{\Alph{appendixc}.\arabic{subappendixc}}
\renewcommand{\thesubsubappendixc}
	{\Alph{appendixc}.\arabic{subappendixc}.\arabic{subsubappendixc}}

\renewcommand{\appendix}[1] {\vspace{12pt}
        \refstepcounter{appendixc}
        \setcounter{figure}{0}
        \setcounter{table}{0}
        \setcounter{lemma}{0}
        \setcounter{theorem}{0}
        \setcounter{corollary}{0}
        \setcounter{definition}{0}
        \setcounter{equation}{0}
        \renewcommand{\thefigure}{\Alph{appendixc}.\arabic{figure}}
        \renewcommand{\thetable}{\Alph{appendixc}.\arabic{table}}
        \renewcommand{\theappendixc}{\Alph{appendixc}}
        \renewcommand{\thelemma}{\Alph{appendixc}.\arabic{lemma}}
        \renewcommand{\thetheorem}{\Alph{appendixc}.\arabic{theorem}}
        \renewcommand{\thedefinition}{\Alph{appendixc}.\arabic{definition}}
        \renewcommand{\thecorollary}{\Alph{appendixc}.\arabic{corollary}}
        \renewcommand{\theequation}{\Alph{appendixc}.\arabic{equation}}
        \noindent{\tenbf Appendix \theappendixc #1}\par\vspace{5pt}}
\newcommand{\subappendix}[1] {\vspace{12pt}
        \refstepcounter{subappendixc}
        \noindent{\bf Appendix \thesubappendixc. {\kern1pt \bfit #1}}
	\par\vspace{5pt}}
\newcommand{\subsubappendix}[1] {\vspace{12pt}
        \refstepcounter{subsubappendixc}
        \noindent{\rm Appendix \thesubsubappendixc. {\kern1pt \tenit #1}}
	\par\vspace{5pt}}

\newcommand{\textlineskip}{\baselineskip=13pt}
\newcommand{\smalllineskip}{\baselineskip=10pt}

 \newcommand{\copyrightheading}[1]
	{\vspace*{-2.5cm}\smalllineskip{\flushleft
	{\footnotesize Presented at DPF-2000}\\
	 }}
\def\abstracts#1#2#3{{
	\centering{\begin{minipage}{4.5in}\baselineskip=10pt\footnotesize
	\parindent=0pt #1\par 
	\parindent=15pt #2\par
	\parindent=15pt #3
	\end{minipage}}\par}} 

\renewenvironment{thebibliography}[1]
	{\frenchspacing
	 \ninerm\baselineskip=11pt
	 \begin{list}{\arabic{enumi}.}
	{\usecounter{enumi}\setlength{\parsep}{0pt}
	 \setlength{\leftmargin 12.7pt}{\rightmargin 0pt} %FOR 1--9 ITEMS
	 \setlength{\itemsep}{0pt} \settowidth
	{\labelwidth}{#1.}\sloppy}}{\end{list}}

\newcounter{itemlistc}
\newcounter{romanlistc}
\newcounter{alphlistc}
\newcounter{arabiclistc}

\newcommand{\fcaption}[1]{
        \refstepcounter{figure}
        \setbox\@tempboxa = \hbox{\footnotesize Fig.~\thefigure. #1}
        \ifdim \wd\@tempboxa > 5in
           {\begin{center}
        \parbox{5in}{\footnotesize\smalllineskip Fig.~\thefigure. #1}
            \end{center}}
        \else
             {\begin{center}
             {\footnotesize Fig.~\thefigure. #1}
              \end{center}}
        \fi}

\newcommand{\tcaption}[1]{
        \refstepcounter{table}
        \setbox\@tempboxa = \hbox{\footnotesize Table~\thetable. #1}
        \ifdim \wd\@tempboxa > 5in
           {\begin{center}
        \parbox{5in}{\footnotesize\smalllineskip Table~\thetable. #1}
            \end{center}}
        \else
             {\begin{center}
             {\footnotesize Table~\thetable. #1}
              \end{center}}
        \fi}

\def\@citex[#1]#2{\if@filesw\immediate\write\@auxout
	{\string\citation{#2}}\fi
\def\@citea{}\@cite{\@for\@citeb:=#2\do
	{\@citea\def\@citea{,}\@ifundefined
	{b@\@citeb}{{\bf ?}\@warning
	{Citation `\@citeb' on page \thepage \space undefined}}
	{\csname b@\@citeb\endcsname}}}{#1}}

\newif\if@cghi
\def\cite{\@cghitrue\@ifnextchar [{\@tempswatrue
	\@citex}{\@tempswafalse\@citex[]}}
\def\citelow{\@cghifalse\@ifnextchar [{\@tempswatrue
	\@citex}{\@tempswafalse\@citex[]}}
\def\@cite#1#2{{$\null^{#1}$\if@tempswa\typeout
	{IJCGA warning: optional citation argument 
	ignored: `#2'} \fi}}

\def\pmb#1{\setbox0=\hbox{#1}
	\kern-.025em\copy0\kern-\wd0
	\kern.05em\copy0\kern-\wd0
	\kern-.025em\raise.0433em\box0}

\def\fnt#1#2{\footnotetext{\kern-.3em
	{$^{\mbox{\scriptsize #1}}$}{#2}}}

\def\fpage#1{\begingroup
\voffset=.3in
\thispagestyle{empty}\begin{table}[b]\centerline{\footnotesize #1}
	\end{table}\endgroup}

\def\runninghead#1#2{\pagestyle{myheadings}
\markboth{{\protect\footnotesize\it{\quad #1}}\hfill}
{\hfill{\protect\footnotesize\it{#2\quad}}}}
\font\tenrm=cmr10
\font\tenit=cmti10 
\font\tenbf=cmbx10
\font\bfit=cmbxti10 at 10pt
\font\ninerm=cmr9

\font\eightrm=cmr8

\def\qed{\hbox{${\vcenter{\vbox{			%HOLLOW SQUARE
   \hrule height 0.4pt\hbox{\vrule width 0.4pt height 6pt
   \kern5pt\vrule width 0.4pt}\hrule height 0.4pt}}}$}}

\begin{document}

\runninghead{Measurement of $d\sigma/dy$ for High Mass Drell-Yan
Pairs $\ldots$} {Measurement of $d\sigma/dy$ for High Mass Drell-Yan
Pairs $\ldots$}

\normalsize\textlineskip
\thispagestyle{empty}
\setcounter{page}{1}

 \copyrightheading{}			%{Vol. 0, No. 0 (1993) 000--000}

\vspace*{0.88truein}

\fpage{1}
\centerline{\bf MEASUREMENT OF $d\sigma/dy$ FOR HIGH MASS}
\vspace*{0.035truein}
\centerline{\bf DRELL-YAN $e^+e^-$ PAIRS AT CDF}
\centerline{\bf }
\vspace*{0.37truein}
\centerline{\footnotesize ARIE BODEK}
\vspace*{0.015truein}
\centerline{\footnotesize\it Department of Physics and Astronomy,
 University of Rochester}
\baselineskip=10pt
\centerline{\footnotesize\it Rochester, NY 14627, USA }
\vspace*{10pt}
\centerline{\footnotesize for the CDF Collaboration (Presented at DPF 2000)}
%\vspace*{0.015truein}
%\centerline{\footnotesize\it Group, Laboratory, Address}
%\baselineskip=10pt
%\centerline{\footnotesize\it City, State ZIP/Zone, Country}
%\vspace*{0.225truein}
%\publisher{(received date)}{(revised date)}

\vspace*{0.21truein}
\abstracts{
We report on the first measurement of the rapidity distribution 
$d\sigma/dy$ over nearly the entire kinematic region of rapidity
for $e^+e^-$ pairs in 
the $Z$-boson region of $66<M_{ee}<116$~GeV$/c^2$ and at higher 
mass  $M_{ee}>116$~GeV$/c^2$. The data sample consists of
108~pb$^{-1}$ of $p\bar{p}$ collisions at $\sqrt{s}=1.8$~TeV
taken by the Collider Detector at Fermilab during 1992--1995.  
The total cross section in the $Z$-boson region is measured to be 
$252 \pm 11$ pb.
The measured total cross section and  $d\sigma/dy$ are compared with 
quantum chromodynamics calculations in leading and higher orders.
% (CDF Note 5422, FERMILAB-CONF-00/249-E, hep-ex/0009067)
}{}{}

\textlineskip			%) USE THIS MEASUREMENT WHEN THERE IS
\vspace*{12pt}			%) NO SECTION HEADING

%\vspace*{1pt}\textlineskip	%) USE THIS MEASUREMENT WHEN THERE IS
%\section{General Appearance}	%) A SECTION HEADING
%\vspace*{-0.5pt}
\noindent

% for references use $^5$''.
Most measurements at high energy proton-antiproton colliders are 
performed in the central rapidity production region, $|y| < 1$. 
A model dependent extrapolation for $|y|>1$ is needed to
extract the total cross section for hard processes such as
top quark production or $W$ and $Z$ boson production.
This extrapolation is made using Monte Carlo programs
(e.g. {\small PYTHIA}, which incorporate
quantum chromodynamics (QCD) calculations in leading order (LO)
or next to leading order (NLO). A previous measurement
of the rapidity distribution, $d\sigma/dy$, for dimuon pairs 
in the $Z$-boson mass region was limited to $|y| < 1$.
In this communication, we present the first measurement of 
$d\sigma/dy$ for $e^+e^-$ pairs in the $Z$-boson mass and high mass
region over nearly the entire kinematic region of rapidity. 
At the Tevatron $p\bar p$ collider, the kinematic limit at the
$Z$-boson mass is $|y|=3.0$, while we measure $|y|$ up to 2.8.
The $d\sigma/dy$ distributions are compared to the predictions of 
QCD in LO and NLO. This measurement is also relevant for precision 
$W$ boson mass measurements at hadron colliders, where $W$'s are 
reconstructed using $e\nu$ and $\mu\nu$ pairs from the Drell-Yan process. 
In hadron-hadron collisions at high energies, massive $e^+e^-$ 
pairs are produced via the Drell-Yan process. 
In the standard model, quark-antiquark annihilation form an 
intermediate $\gamma^*$ or $Z$ ($\gamma^*/Z$) vector boson, 
which then decays into an $e^+e^-$ pair. In LO, 
the momentum fraction $x_1$ ($x_2$) of the partons in the 
proton (antiproton) are related to the rapidity, 
$y$, of the boson via the equation $ x_{1,2} = ({M}/{\sqrt{s}}) e^{\pm y}$.
Here $s$ is the center of mass energy, and $M$ is the mass of the 
dilepton pair. Therefore, dilepton pairs which are produced at large
rapidity originate from events in which one parton is
at large $x$ and another parton is at very small $x$.
Since the quark distributions for $x$ up to 0.9 are well constrained by the 
deep-inelastic lepton scattering experiments,
comparisons of data and theory for $d\sigma/dy$, and the total
cross sections provide a test of the theory, \mbox{e.g.}, missing 
NNLO$^1$ or power correction terms.
The $e^+e^-$ pairs are from 108~pb$^{-1}$ of $p\bar{p}$ collisions
at $\sqrt{s}=1.8$~TeV taken by the Collider Detector at Fermilab
(CDF) during 1992--1993 ($18.7 \pm 0.7$~pb$^{-1}$) and 1994--1995 
($89.1 \pm 3.7$~pb$^{-1}$).
    In a previous letter, we presented $d\sigma/dP_{\rm T}$ of $Z$
    boson using has three categories of $e^+e^-$ pairs:
    central-central (CC), central-end plug (CP), and central-forward (CF).
    This analysis extends the sample to the forward rapidity region by 
    including plug-plug (PP) and plug-forward (PF) events. The
    inclusion of these events increases the event sample by $20\%$ and
    allows for measurement of $Z$ bosons with $|y|$ up to 2.8.
    An improvement in this analysis is the additional VTX tracking
    requirements for plug and forward electrons.
    The VTX covers the entire rapidity range in this study, and plays
    an important role in removing background in the high $\eta$ region
    which is not covered by the CTC.

The data sample is divided into two mass bins: 
the $Z$ region ($66<M_{ee}<116$~GeV/$c^2$) and the high mass
region ($M_{ee}>116$~GeV/$c^2$). After all cuts, the numbers of CC, CP, 
CF, PP, and PF events in the $Z$ mass region are 2894, 3811, 621, 
1236, and 589, respectively. The backgrounds are low and are 
estimated using the data.
The PP+PF events extend the acceptance of the $y$ 
measurement to $|y|=2.8$, and significantly increase our statistics by 
extending the acceptance beyond $y > 1.2$.

Figures~1(a) and 1(b) compare the measured 
$d\sigma/dy$'s to theoretical predictions in the $Z$ mass 
and high mass regions, respectively. The top horizontal axes on
these figures are the corresponding values of the $x_1$ and $x_2$ 
as a function of $y$. The predictions are LO calculations with 
CTEQ5L PDFs and NLO calculations with MRST99 
and CTEQ5M-1 PDFs.
The predictions in Figure~1(a) have been normalized by a 
factor ``F'', the ratio of measured total cross section to 
the prediction (F=1.51, 1.14, and 1.13 for the CTEQ5L,
MRST99 and CTEQ5M-1 PDFs, respectively).
The predictions in Figure~1(b) are normalized to the 
data using the factor F from the $Z$ mass region. 
As the $\chi^2$ values listed in Figure~1(a) indicate,
the LO calculation using recent LO PDFs does not fit the shape
as well as the NLO calculation with the most recent NLO PDFs.

%\begin{figure}[htbp]
%\vspace*{13pt}
%\centerline{\vbox{\hrule width 5cm height0.001pt}}
%\vspace*{1.4truein}		%ORIGINAL SIZE=1.6TRUEIN x 100% - 0.2TRUEIN
%\centerline{\vbox{\hrule width 5cm height0.001pt}}
%\vspace*{13pt}
%\fcaption{Labeled tree {\footnotesize\it T}.}
%\end{figure}

Model independent measurements of the total production cross sections 
for  $e^+e^-$ pairs are extracted by integrating the measured values 
of $d\sigma/dy$. The extracted cross section in the $Z$ mass region is 
$252.1 \pm 3.9\:{\rm (stat.)} \pm 1.6\:{\rm (syst.)} \pm 9.8\:{\rm (lum.)}$~pb.
The corresponding $\sigma(p\bar p\rightarrow Z) \cdot Br(Z\rightarrow
ee)$ is $253 \pm 4({\rm stat.+syst.}) \pm 10 {\rm (lum.)}$~pb.
The total cross section measurements can also be compared to QCD calculations.
Fixed order QCD calculations have uncertainties from PDF measurements and
corrections from higher orders of perturbation theory,
\mbox{i.e.}, the $K$-factor.
The NLO-to-LO total cross section correction is significant: $K \sim 1.4$.
In contrast, the NLO total cross section is lower than NNLO
prediction by only $2.3\%$. The NNLO prediction
with the latest NLO MRST99 PDFs is $227 \pm 9$~pb,
where the $4\%$ error is mostly from uncertainties in the NLO PDFs.
Although a full set of NNLO PDFs is not available, recent 
estimates
of NNLO PDFs indicate$^1$ that the NNLO PDFs will increase the theoretical
cross sections by $1-4\%$ (e.g. MRST00(NNLO) PDF yields $230 \pm 9$~pb)
Given these uncertainties, the theoretical expectation is consistent with
the $Z$ cross section measurements.
The measurement of the Drell-Yan total cross section in the high mass
region is $4.0 \pm 0.4\:{\rm (stat.+syst.)} \pm 0.2\:{\rm (lum.)}$~pb.
The corresponding prediction of the total cross section from the NNLO QCD 
theory using MRST99 PDFs is $3.3$ pb. Additional details on this work can 
be found in Reference 2.  

\nonumsection{References}

\begin{figure}[htbp]
%\vspace*{13pt}
%\begin{center}
\centerline{\mbox{\epsfxsize=3.37in \epsfysize=2.25in \epsffile{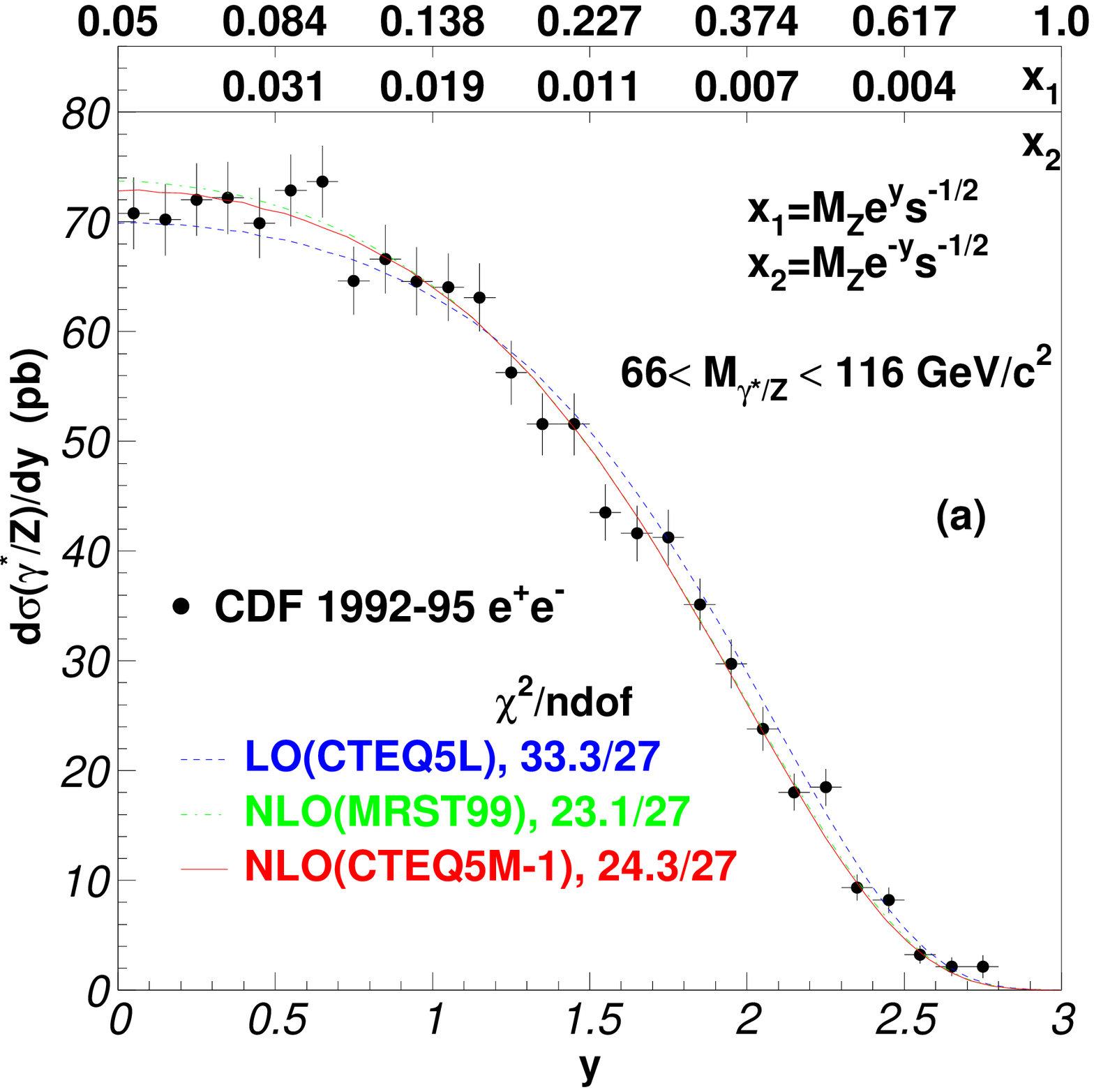}}}
% PRL 2 column size
%\mbox{\epsfxsize=6.00in \epsffile{dsdy.eps}}   % Preprint size
\centerline{\mbox{\epsfxsize=3.37in \epsfysize=2.25in \epsffile{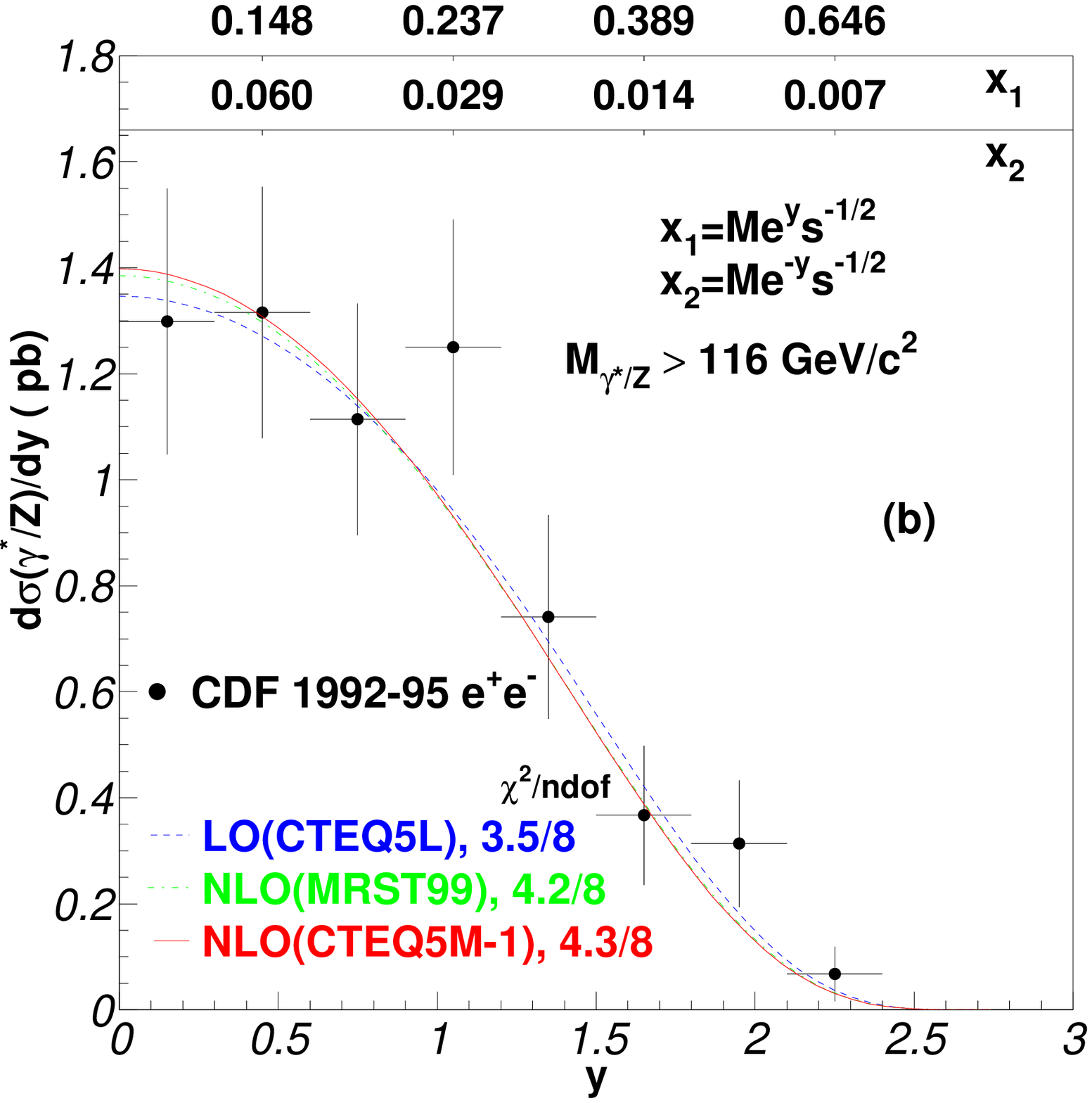}}}
% PRL 2 column size
%\end{center}
\fcaption{$d\sigma/dy$ 
distribution of $e^+e^-$ pairs: (a) in the $Z$ boson mass 
($66<M_{ee}<116$~GeV/$c^2$) region. 
(b) in the high mass ($M_{ee}>116$~GeV/$c^2$) region. The $M$ used to
obtain $x_1$ and $x_2$ in (b) is the mean mass over the bin. The error
bars on the data include statistical errors only. The
theoretical predictions have been normalized to the data in the
$Z$ boson mass region.}
%\label{fig1}
\end{figure}
\end{document}